\documentclass[12pt,a4paper]{elsart}
\usepackage{amssymb,amsmath,stmaryrd}
\usepackage[mathscr]{eucal}
\usepackage{pstricks,pst-node,pst-text,pst-3d}
\usepackage{lineno}

\newtheorem{theorem}{Theorem}
\newtheorem{proposition}{Proposition}

\newtheorem{corollary}{Corollary}

\renewcommand{\labelenumi}{(\roman{enumi})}
\def \G {\mathcal{G}}
\def \Gk {\mathcal{G}^k}
\def \MGk {\mathcal{MG}^k}
\def \IG {\mathcal{G}_\infty}
\def \IGk {\mathcal{G}_\infty^k}
\def \PkN {\mathcal{P}^k(N)}
\def \PkNe {\mathcal{P}^k_*(N)}
\def \PkLe {\mathcal{P}^k_*(L)}
\def \AB {\mathcal{A}(B)}
\def \core {\mathcal{C}}
\def \corek {\mathcal{C}^k}
\def \mcorek {\mathcal{MC}^k}

\newcounter{remark}
\newenvironment{remark}
{\begin{quote}\textsc{Remark} \stepcounter{remark} \arabic{remark}:}
{\end{quote}}
\newcounter{example}
\newenvironment{example}
{\begin{quote}\textsc{Example} \stepcounter{example} \arabic{example}:}
{\end{quote}}

\newenvironment{proof}{\medskip\noindent \bf Proof: \rm}{\hspace*{\fill}
$\blacksquare$ \newline \medskip}

\begin{document}
\begin{frontmatter}

\title{On the vertices of the $k$-additive core} 

\author[p1]{Michel GRABISCH\corauthref{cor}},
\corauth[cor]{Corresponding author. Tel (+33) 1-44-07-82-85, Fax (+33)
1-44-07-83-01.}
\ead{michel.grabisch@univ-paris1.fr}
\author[mad]{Pedro MIRANDA}
\ead{pmiranda@mat.ucm.es}

\address[p1]{University of Paris I
-- Panth\'eon-Sorbonne, Centre d'\'Economie de la Sorbonne, 106-112 Bd de
l'H\^opital, 75013 Paris, France}
\address[mad]{Universidad Complutense de Madrid, Plaza de Ciencias, 3, 28040
Madrid, Spain}

\date{\today}

\maketitle

\begin{abstract}
The core of a game $v$ on $N$, which is the set of additive games $\phi$
dominating $v$ such that $\phi(N)=v(N)$, is a central notion in cooperative game
theory, decision making and in combinatorics, where it is related to submodular
functions, matroids and the greedy algorithm. In many cases however, the core is
empty, and alternative solutions have to be found. We define the $k$-additive
core by replacing additive games by $k$-additive games in the definition of the
core, where $k$-additive games are those games whose M\"obius transform vanishes
for subsets of more than $k$ elements. For a sufficiently high value of $k$, the
$k$-additive core is nonempty, and is a convex closed polyhedron. Our aim is to
establish results similar to the classical results of Shapley and Ichiishi on
the core of convex games (corresponds to Edmonds' theorem for the greedy
algorithm), which characterize the vertices of the core.
\end{abstract}

\end{frontmatter}

\linenumbers
\section{Introduction}
Given a finite set $N$ of $n$ elements, and a set function $v:2^N\rightarrow
\mathbb{R}$ vanishing on the empty set (called hereafter a \emph{game}), its
\emph{core} $\core(v)$ is the set of additive set functions $\phi$ on $N$ such
that $\phi(S)\geq v(S)$ for every $S\subseteq N$, and $\phi(N)=v(N)$. Whenever
nonempty, the core is a convex closed bounded polyhedron.

In many fields, the core is a central notion which has deserved a lot of
studies. In cooperative game theory, it is the set of imputations for players so
that no subcoalition has interest to form \cite{vnm44}. In decision making under
uncertainty, where games are replaced by \emph{capacities} (monotonic games), it
is the set of probability measures which are coherent with the given
representation of uncertainty \cite{wal91}. More on a combinatorial point of
view, cores of convex games are also known as base polytopes associated to
supermodular functions \cite{lov83,fuj91}, for which the greedy algorithm is
known to be a fundamental optimization technique. Many studies have been done
along this line, e.g., by Faigle and Kern for the matching games
\cite{fakefeho98}, and cost games \cite{fake00}. In game theory, which will be
our main framework here, related notions are the selectope \cite{bijilepe00},
and the Shapley value with many of its variations on combinatorial structures (see, e.g.,
\cite{albibrji04}).

It is a well known fact that the core is nonempty if and only if the game is
balanced \cite{bon63}. In the case of emptiness, an alternative solution has to
be found. One possibility is to search for games more general than additive
ones, for example $k$-additive games and capacities proposed by Grabisch
\cite{gra96f}. In short, $k$-additive games have their M\"obius transform
vanishing for subsets of more than $k$ elements, so that 1-additive games are
just usual additive games. Since any game is a $k$-additive game for some $k$
(possibly $k=n$), the $k$-additive core, i.e., the set of dominating
$k$-additive games, is never empty provided $k$ is high enough. The authors have
justified this definition in the framework of cooperative game theory
\cite{migr04}. Briefly speaking, an element of the $k$-additive core implicitely
defines by its M\"obius transform an imputation (possibly negative), which is
now defined on groups of at most $k$ players, and no more on individuals. By
definition of the $k$-additive core, the total worth assigned to a coalition
will be always greater or equal to the worth the coalition can achieve by
itself; however, the precise sharing among players has still to be decided
(e.g., by some bargaining process) among each group of at most $k$ players.

In game theory, elements of the core are imputations
for players, and thus it is natural that they fulfill monotonicity. We call
monotonic core the core restricted to monotonic games (capacities). We will see
in the sequel that the core is usually unbounded, while the monotonic core is
not.  

The properties of the (classical) core are well known, most remarkable being the
result characterizing the core of convex games, where the set of vertices is
exactly the set of additive games induced by maximal chains (or equivalently by
permutations on $N$) in the Boolean lattice $(2^N,\subseteq)$. This has been shown by
Shapley \cite{sha71}, and later Ichiishi proved the converse implication
\cite{ich81}. This result is also known in the field of matroids, since vertices
of the base poytope can be found by a greedy algorithm. 

A natural question arises: is it possible to generalize the Shapley-Ichiishi
theorem for $k$-additive (monotonic) cores? More precisely, can we find the set
of vertices for some special classes of games? Are they induced by some
generalization of maximal chains? The paper shows that the answer is more
complex than expected. It is possible to define notions similar to permutations
and maximal chains, so as to generate vertices of the $k$-additive core of
$(k+1)$-monotone games, a result which is a true generalization of the
Shapley-Ichiishi theorem, but this does not permit to find all vertices of the
core. A full analytical description of vertices seems to be difficult to find,
but we completely explicit the case $k=n-1$.

After a preliminary section introducing necessary concepts, Section
\ref{sec:orde} presents our basic ingredients, that is, orders on subsets of at
most $k$ elements, and achievable families, which play the role of maximal
chains in the classical case. Then Section \ref{sec:vert} presents the main
result on the characterization of vertices for ($k+1$)-monotone games induced by
achievable families. 

\section{Preliminaries}\label{sec:prel}
Throughout the paper, $N:=\{1,\ldots,n\}$ denotes a set of $n$ elements
(players in a game, nodes of a graph, etc.). We
use indifferently $2^N$ or $\mathcal{P}(N)$ for denoting the set of subsets of $N$,
and the set of subsets of $N$ containing at most $k$ elements is denoted by
$\PkN$, while $\PkNe:=\PkN\setminus\{\emptyset\}$. For convenience, subsets like
$\{i\}, \{i,j\}, \{2\}, \{2,3\},\ldots$ are written in the compact form $i, ij,
2, 23$ and so on. 

A \emph{game} on $N$ is a function $v:2^N\rightarrow \mathbb{R}$ such that
$v(\emptyset)=0$. The set of games on $N$ is denoted by $\mathcal{G}(N)$. For
any $A\in 2^N\setminus\{\emptyset\}$, the \emph{unanimity game centered on $A$}
is defined by $u_A(B):=1$ iff $B\supseteq A$, and 0 otherwise.

A game $v$ on $N$ is said to be:
\begin{enumerate}
\item \emph{additive} if $v(A\cup B)=v(A)+v(B)$ whenever $A\cap
B=\emptyset$;
\item \emph{convex} if $v(A\cup B)+v(A\cap B)\geq v(A)+v(B)$, for all
  $A,B\subseteq N$;
\item \emph{monotone} if $v(A)\leq v(B)$ whenever $A\subseteq B$;
\item \emph{$k$-monotone} for $k\geq 2$ if for any family of $k$ subsets
  $A_1,\ldots A_k$, it holds
\[
v(\bigcup_{i=1}^k A_i) \geq \sum_{\substack{K\subseteq \{1,\ldots,k\}\\
  K\neq\emptyset}} (-1)^{|K|+1}v(\bigcap_{j\in K}A_j)
\]
\item \emph{infinitely monotone} if it is $k$-monotone for all $k\geq 2$.
\end{enumerate}

Convexity corresponds to 2-monotonicity. Note that $k$-monotonicity implies
$k'$-monotonicity for all $2\leq k'\leq k$. Also, $(n-2)$-monotone games on $N$
are infinitely monotone \cite{bar00}. The set of monotone games on $N$ is
denoted by $\mathcal{MG}(N)$, while the set of infinitely monotone games
is denoted by $\IG(N)$.

Let $v$ be a game on $N$. The \emph{M\"obius transform} of $v$ \cite{rot64}
(also called \emph{dividends} of $v$, see Harsanyi \cite{har63}) is a function
$m:2^N\to\mathbb{R}$ defined by:
\[
m(A) := \sum_{B\subseteq A} (-1)^{|A\setminus B|}v(B),\quad \forall A\subseteq N.
\]
The M\"obius transform is invertible since one can recover $v$ from $m$ by:
\[
v(A) = \sum_{B\subseteq A}m(B),\quad\forall A\subseteq N.
\]
If $v$ is an additive game, then $m$ is non null only for singletons, and
$m(\{i\})=v(\{i\})$. The M\"obius transform of $u_A$ is given by $m(A)=1$ and
$m$ is 0 otherwise.

A game $v$ is said to be {\em $k$-additive} \cite{gra96f}
for some integer $k\in\{1,\ldots,n\}$ if $m(A)=0$ whenever $|A|>k$, and there
exists some $A$ such that $|A|=k$, and $m(A)\neq 0$.

Clearly, 1-additive games are additive. The set of games on $N$ being at most
$k$-additive (resp. infinitely monotone games at most $k$-additive) is denoted
by $\Gk(N)$ (resp. $\IGk(N)$). As above, replace $\mathcal{G}$ by $\mathcal{MG}$
if monotone games are considered instead. 

We recall the fundamental following result.
\begin{proposition}\label{prop:chj}
\cite{chja89} Let $v$ be a game on $N$. For any $A,B\subseteq N$, with
$A\subseteq B$, we denote  $[A,B]:=\{L\subseteq N\mid A\subseteq L\subseteq B\}$.
\begin{enumerate}
\item Monotonicity is equivalent to 
\[
\sum_{L\in[i,B]}m(L)\geq 0,\quad \forall B\subseteq N, \quad \forall i\in B.
\]
\item For $2\leq k\leq n$, $k$-monotonicity is equivalent to
\[
\sum_{L\in[A,B]}m(L)\geq 0,\quad \forall A, B\subseteq N, A\subseteq B,\quad
2\leq |A|\leq k.
\]
\end{enumerate}
\end{proposition}
Clearly, a monotone and infinitely monotone game has a nonnegative M\"obius
transform.

\medskip

The \emph{core} of a game $v$ is defined by:
\[
\core(v):=\{\phi\in\G^1(N)\mid \phi(A)\geq v(A),\quad\forall A\subseteq N,
\text{ and }\phi(N)=v(N)\}.
\]

A \emph{maximal chain} in $2^N$ is a sequence of subsets
$A_0:=\emptyset,A_1,\ldots,A_{n-1},A_n:=N$ such that $A_i\subset
A_{i+1}$, $i=0,\ldots,n-1$. The set of maximal chains of $2^N$ is
denoted by $\mathcal{M}(2^N)$.

To each maximal chain $C:=\{\emptyset,A_1,\ldots,A_n=N\}$ in
$\mathcal{M}(2^N)$ corresponds a unique permutation $\sigma$ on
$N$ such that $A_1=\sigma(1)$, $A_2\setminus A_1=\sigma(2)$,
\ldots, $A_n\setminus A_{n-1}=\sigma(n)$. The set of all
permutations over $N$ is denoted by $\mathfrak{S}(N)$. Let $v$
be a game.  Each permutation $\sigma$ (or maximal chain $C$)
induces an additive game $\phi^\sigma$ (or $\phi^C$) on $N$ defined by:
\[
\phi^\sigma(\sigma(i)):=v(\{\sigma(1),\ldots,\sigma(i)\})-v(\{\sigma(1),\ldots,\sigma(i-1)\})
\]
or
\[
\phi^C(\sigma(i)):=v(A_i)-v(A_{i-1}),\quad \forall i\in N.
\]
with the above notation. The following is immediate.
\begin{proposition}\label{prop:mave}
Let $v$ be a game on $N$, and $C$ a maximal chain of $2^N$. Then
\[
\phi^C(A) = v(A),\quad\forall A\in C.
\]
\end{proposition}

\begin{theorem}\label{th:sha}
The following propositions are equivalent.
\begin{enumerate}
\item $v$ is a convex game.
\item All additive games $\phi^\sigma$, $\sigma\in\mathfrak{S}(N)$, belong
to the core of $v$.
\item $\core(v) = \mathrm{co}(\{\phi^\sigma\}_{\sigma\in\mathfrak{S}(N)})$.
\item $\mathrm{ext}(\core(v)) = \{\phi^\sigma\}_{\sigma\in\mathfrak{S}(N)}$,
\end{enumerate}
where $ \mathrm{co}(\cdot)$ and $\mathrm{ext}(\cdot)$ denote respectively the convex
hull of some set, and the extreme points of some convex set.
\end{theorem}
(i) $\Rightarrow$ (ii) and (i) $\Rightarrow$ (iv) are due to Shapley
\cite{sha71}, while (ii) $\Rightarrow$ (i) was proved by Ichiishi \cite{ich81}.

A natural extension of the definition of the core is the following. For some
integer $1\leq k\leq n$, the \emph{$k$-additive core} of a game $v$ is defined by:
\[
\corek(v):=\{\phi\in\Gk(N)\mid \phi(A)\geq v(A),\quad \forall A\subseteq N,
\phi(N)=v(N)\}. 
\]
In a context of game theory where elements of the core are imputations, it is
natural to consider that monotonicity must hold, i.e.,  the
imputation allocated to some coalition $A\in\PkNe$ is larger than for any subset
of $A$. We call it the \emph{monotone $k$-additive core}: 
\[
\mcorek(v):=\{\phi\in\MGk(N)\mid \phi(A)\geq v(A),\quad \forall A\subseteq N,
\phi(N)=v(N)\}. 
\]
We introduce also the \emph{core of $k$-additive infinitely monotone games}:
\[
\corek_\infty(v):=\{\phi\in\IGk(N)\mid \phi(A)\geq v(A),\quad \forall A\subseteq
N, \text{ and }\phi(N)=v(N)\}.
\]
The latter is introduced just for mathematical convenience, and has no clear
application. Note that $\core(v)=\core^1(v)=\core^1_\infty(v)$.

\section{Orders on $\PkNe$ and achievable families}\label{sec:orde}
As our aim is to give a generalization of the Shapley-Ichiishi results, we need
counterparts of permutations and maximal chains. These are given in this
section. Exact connections between our material and permutations and maximal
chains will be explicited at the end of this section. First, we introduce total
orders on subsets of at most $k$ elements as a generalization of permutations.

We denote by $\prec$ a total (strict) order on $\PkNe$, $\preceq$ denoting the
corresponding weak order. 
\begin{enumerate}
\item $\prec$ is said to be \emph{compatible} if for all $A,B\in\PkNe$, $A\prec
  B$ if and only if $A\cup C\prec B\cup C$ for all $C\subseteq N$ such that
  $A\cup C, B\cup C\in\PkNe$, $A\cap C=B\cap C=\emptyset$.
\item $\prec$ is said to be
\emph{$\subseteq$-compatible} if $A\subset B$ implies $A\prec B$. 
\item $\prec$ is said to be \emph{strongly compatible} if it is compatible and
$\subseteq$-compatible.
\end{enumerate}

We introduce the \emph{binary order} $\prec^2$ on $2^N$ as follows. To any
subset $A\subseteq N$ we associate an integer $\eta(A)$, whose binary code is
the indicator function of $A$, i.e., the $i$th bit of $\eta(A)$ is 1 if $i\in
A$, and 0 otherwise. For example, with $n=5$, $\{1,3\}$ and $\{4\}$ have
binary codes $00101$ and $01000$ respectively, hence $\eta(\{1,3\})=5$ and
$\eta(\{4\}) = 8$. Then $A\prec^2 B$ if $\eta(A)<\eta(B)$. This gives
\begin{multline}\label{eq:bino}
1\prec^2 2\prec^2 12\prec^2 3\prec^2 13\prec^2 23\prec^2 123\prec^2 4\prec^2
14\prec^2 24\prec^2 \\ 
124\prec^2 34\prec^2 134\prec^2 234\prec^2 1234\prec^2 5\prec^2 \ldots 
\end{multline}
Note the recursive nature of $\prec^2$. Obviously, $\prec^2$ is a strongly
compatible order, as well as all its restrictions to $\PkNe$, $k=1,\ldots,n-1$.

\medskip

We introduce now a generalization of maximal chains associated to permutations. 
Let $\prec$ be a total order on $\PkNe$. For any $B\in \PkNe$, we define
\[
\AB:=\{A\subseteq N\mid [A\supseteq B] \text{ and } [\forall K\subseteq A\text{
    s.t. } K\in\PkNe, \text{ it holds }
K\preceq B]\} 
\]
the \emph{achievable family of $B$}.
\begin{example}
Consider $n=3$, $k=2$, and the following order: $1\prec 2\prec 12\prec 13\prec
23\prec 3$. Then
\begin{align*}
&\mathcal{A}(1)=\{1\},\quad
\mathcal{A}(2)=\{2\},\quad\mathcal{A}(12)=\{12\},\quad\mathcal{A}(13)=\mathcal{A}(23)=\emptyset,\\
& \mathcal{A}(3)=\{3,13,23,123\}.
\end{align*}
\end{example} 
\begin{proposition}\label{prop:0}
$\{\AB\}_{B\in\PkNe}$ is a partition of $\mathcal{P}(N)\setminus \{\emptyset\}$.
\end{proposition}
\begin{proof}
  Let $\emptyset\neq C\in\mathcal{P}(N)$. It suffices to show that there is a
  unique $B\in\PkNe$ such that $C\in\AB$. Let $K_1,K_2,\ldots,K_p$ be the
  nonempty collection of subsets of $C$ in $\PkN$, assuming $K_1\prec
  K_2\prec\cdots\prec K_p$. Then $C\in\mathcal{A}(K_p)$ is the unique
  possibility, since any $B$ outside the collection will fail to satisfy the
  condition $B\subseteq C$, and any $B\neq K_p$ inside the collection will fail
  to satisfy the condition $K_p\preceq B$.
\end{proof}

\begin{proposition}\label{prop:1}
  For any $B\in\PkNe$ such that $\AB\neq\emptyset$, $(\AB,\subseteq)$
  is an inf-semilattice, with bottom element $B$.
\end{proposition}
\begin{proof}
If $\AB\neq\emptyset$, any $C\in\AB$ contains $B$, hence every
$K\subseteq B\subseteq C$, $K\in\PkNe$, is
such that $K\preceq B$. Hence $B\in\AB$, and it is the smallest element.

Let $K,K'\in\AB$, assuming $\AB$ contains at least 2 elements (otherwise, we are
done). $K\in\AB$ is equivalent to $K\supseteq B$ and any $L\subseteq K$,
$L\in\PkNe$ is such that $L\preceq B$. The same holds for $K'$. Therefore,
$K\cap K'\supseteq B$, and if $L\subseteq K\cap K'$, $L\in\PkNe$, then
$L\subseteq K$ and $L\subseteq K'$, which entails $L\preceq B$. Hence $K\cap
K'\in\AB$.
\end{proof}

From the above proposition we deduce:
\begin{corollary}\label{cor:1}
  Let $B\in\PkNe$ and $\prec$ be some total order on $\PkNe$. Then
  $\AB\neq\emptyset$ if and only if for all $C\in\PkNe$, $C\subseteq B$ implies
  $C\preceq B$.  Consequently, if $|B|=1$ then $\AB\neq \emptyset$.
\end{corollary}
\begin{corollary}\label{lem:2}
$\AB\neq\emptyset$ for all $B\in\PkNe$ if and only if $\prec$ is
  $\subseteq$-compatible. 
\end{corollary}

It is easy to build examples where achievable families are not lattices. 
\begin{example}
Consider $n=4, k=2$ and the following order: 2, 3, 24, 12, 4, 13, 34, 1, 23,
14. Then $\mathcal{A}(23)=\{23, 123, 234\}$, and $1234\not\in \mathcal{A}(23)$
since $14\succ 23$.
\end{example}

Assuming $\AB$ is a lattice, we denote by $\check{B}$ its top element.
\begin{proposition}\label{prop:3}
Let $\prec$ be a total order on $\PkNe$. Consider $B\in\PkNe$ such that $\AB$ is
a lattice. Then it is a Boolean lattice isomorphic to
$(\mathcal{P}(\check{B}\setminus B), \subseteq)$. 
\end{proposition}
\begin{proof}
  It suffices to show that $\AB=\{B\cup K\mid K\subseteq \check{B}\setminus
  B\}$. Taking $\check{K}:=\check{B}\setminus B$, we have $B\cup \check{K}\in \AB$.
  Hence, any $L\subseteq B\cup \check{K}$, $L\in\PkNe$, is such that $L\preceq
  B$. This is a fortiori true for $L\subseteq B\cup K$, $L\in\PkNe$, $\forall
  K\subseteq \check{K}$. Hence $B\cup K$ belongs to $\AB$, for all $K\subseteq
  \check{K}$. 
\end{proof}

\begin{proposition}\label{prop:2}
Assume $\prec$ is compatible. For any $B\in\PkNe$ such that $\AB\neq\emptyset$,
$\AB$ is the Boolean lattice $[B,\check{B}]$.
\end{proposition}
\begin{proof}
If $\AB$ is a lattice, we know by Prop. \ref{prop:3} that it is a Boolean
lattice with bottom element $B$. Since we know that $\AB$ is an inf-semilattice
by Prop. \ref{prop:1}, it remains to show that $K,K'\in\AB$ implies $K\cup K'\in
\AB$. Assume $K\cup K'\not\in\AB$. Then there exists $L\subseteq K\cup K'$,
$L\in\PkNe$ such that $L\succ B$. Necessarily, $L\setminus K\neq\emptyset$,
otherwise $L\subseteq K$ and $K\in \AB$ imply $L\prec B$, a
contradiction. Similarly, $L\setminus K'\neq\emptyset$. Moreover,
$L\not\subseteq B$ since $\AB\neq\emptyset$ (see Cor. \ref{cor:1}).

We consider $D:=L\setminus K$, not empty by definition of $L$. Since $L\setminus
D\subseteq K$ and $L\setminus D\in\PkNe$, we have $L\setminus D\preceq B$, with
strict inequality since $L\setminus D$ has elements outside $K\cap K'$, hence
outside $B$ (see Figure below).
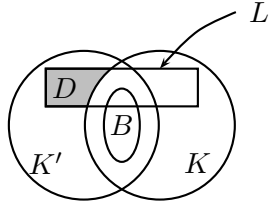
\begin{center}
\psset{unit=0.5cm}
\pspicture(0,0)(6,5)
\pspolygon[fillstyle=solid,fillcolor=lightgray](1,2.5)(1,3.5)(5,3.5)(5,2.5)
\pscircle[fillstyle=solid](4,2){2}
\pspolygon(1,2.5)(1,3.5)(5,3.5)(5,2.5)
\pscircle(2,2){2}
\psellipse(3,2)(0.5,1)
\rput(1,1){\small $K'$}
\rput(5,1){\small $K$}
\rput(3,2){\small $B$}
\rput(1.5,3){\small $D$}
\pscurve{<-}(4,3.5)(5,4.5)(6,5)
\uput[0](6,5){\small $L$}
\endpspicture
\end{center}

Suppose first that $|B|<k$, and let $D:=\{i,j,\ldots\}$. We have $B\cup
l\in\PkNe$ and $B\cup l\subseteq K'$ for any $l\in D$, which implies $B\cup
l\prec B$. Taking $l=i$, by compatibility, $L\setminus D\prec B$ implies
$(L\setminus D)\cup i\prec B\cup i\prec B$. By compatibility again, $(L\setminus
D)\cup i\prec B$ implies $(L\setminus D)\cup i\cup j\prec B\cup j\prec
B$. Continuing the process till all elements of $D$ have been taken, we finally
end with $L\prec B$, a contradiction. 

Secondly, assume that $|B|=k$. Take $K''\subset B$ such that
$K''\supseteq L\cap B$ and $|K''\cup D|=k$, which is always possible by
assumption. Since $K''\subset B\subseteq K$ and $K''\in\PkNe$, we have $K''\prec
B$. Then 
\begin{enumerate}
\item Either $L\setminus D\prec K''\prec B$. By compatibility, $L\setminus
  D\prec K''$ implies $L\prec K''\cup D$. Since $K''\cup D\in\PkNe$ and $K''\cup
  D\subseteq K'$, we deduce that $K''\cup D\prec B$, hence $L\prec B$, a
  contradiction.
\item Or $K''\prec L\setminus D\prec B$. Since $(L\setminus D)\cap(B\setminus
  K'')=\emptyset$, from compatibility $K''\prec L\setminus D$ implies
  $B=K''\cup(B\setminus K'')\prec (L\setminus D)\cup(B\setminus K'')$. We have
  by assumption $|(L\setminus D)\cup(B\setminus K'')|=|L|\leq k$ and
  $(L\setminus D)\cup(B\setminus K'')\subseteq K$, from which we deduce
  $(L\setminus D)\cup(B\setminus K'')\prec B$. Hence we get $B\prec B$, a
  contradiction.
\end{enumerate}
\end{proof}

The following example shows that compatibility is not a necessary condition. 
\begin{example}
Consider $n=4$, $k=2$, and the following order: 1, 3, 2, 12, 23, 13, 4, 14, 24,
34. This order is not compatible since $3\prec 2$ and $12\prec 13$. We obtain:
\begin{gather*}
  \mathcal{A}(1)=1, \quad \mathcal{A}(3)=3, \quad \mathcal{A}(2)=2, \quad
  \mathcal{A}(12)=12, \quad \mathcal{A}(23)=23, \quad
  \mathcal{A}(13)=\{13,123\},\\ \mathcal{A}(4)=4, \quad \mathcal{A}(14)=14,
  \quad \mathcal{A}(24)=\{24,124\}, \quad\mathcal{A}(34)=\{34,134,234,1234\}.
\end{gather*}
All families are lattices.
\end{example}
In the above example, $\prec$ was $\subseteq$-compatible. However, this is not
enough to ensure that achievable families are lattices, as shown by the
following example.
\begin{example}
Let us consider the following $\subseteq$-compatible order with $n=4$ and $k=2$:
\[
3\prec 4\prec34\prec 2 \prec 24 \prec 1 \prec 13\prec 12\prec 23\prec 14.
\]
Then $\mathcal{A}(23)=\{23, 123, 234\}$.
\end{example}

We give some fundamental properties of achievable families when they are
lattices, in particular of their top elements. 
\begin{proposition}\label{prop:4}
  Assume $\prec$ is compatible, and consider a nonempty achievable family $\AB$,
  with top element $\check{B}$. Then $\{\mathcal{A}(B_i)\mid B_i\in\PkNe,
  B_i\subseteq \check{B},\mathcal{A}(B_i)\neq \emptyset \}$ is a partition of
  $\mathcal{P}(\check{B})\setminus \{\emptyset\}$.
\end{proposition}
\begin{proof}
  We know by Prop. \ref{prop:0} that all $\mathcal{A}(B_i)$'s are disjoint. It
  remains to show that (1) any $K\subseteq \check{B}$ is in some
  $\mathcal{A}(B_i)$, $B_i\subseteq \check{B}$, and (2) conversely that any
  $K$ in such $\mathcal{A}(B_i)$ is a subset of $\check{B}$.

  (1) Assume $K\in\mathcal{A}(B_i)$, $B_i\not\subseteq \check{B}$. Then
  $B_i\subseteq K\subseteq \check{B}$, a contradiction.

(2) Assume $K\in\mathcal{A}(B_i)$, $B_i\subseteq \check{B}$, and $K\not\subseteq
\check{B}$. Then there exists $l\in K$ such that $l\not\in \check{B}$ (and hence
not in $B_i$). Note that this implies $B_i\cup l\prec B_i$, provided
$|B_i|<k$. First we show that $l\prec j$ for any $j\in B_i$. Since $K\supseteq
B_i\cup\{l\}$, we deduce that for any $j\in B_i$, $\{j,l\}\prec B_i$ and $l\prec
B_i$. If $B_i=\{j\}$, we can further deduce that $l\prec j$. Otherwise, if
$B_i=\{j,j'\}$, from $\{j,l\}\prec \{j,j'\}$ and $\{j',l\}\prec\{j,j'\}$, by
compatibility $l\prec j$ and $l\prec j'$. Generalizing the above, we conclude
that $l\prec j$ for all $j\in B_i$.

Next, if $l\not\in \check{B}$, it means that for some $B'\subseteq \check{B}$
such that $B'\cup l\in\PkNe$, we have $B'\cup l\succ B$
(otherwise $l$ should belong to $\check{B}$). We prove that $B'\not\supseteq
B_i$. The case $|B_i|=k$ is obvious, let us consider $|B_i|<k$. Suppose on the contrary that $B'=B_i\cup L$, $L\subseteq N\setminus
B_i$. Then $B_i\cup l\prec B_i$ implies that $B'\cup l=B_i\cup l\cup L\prec
B_i\cup L\prec B$, the last inequality coming from $B_i\cup L\subseteq
\check{B}$, $B_i\cup L\in\PkNe$. But $B'\cup l\succ B$, a contradiction.  

Choose any $j\in B_i\setminus B'$. Since $j\succ l$,
we deduce $B'\cup j\succ B'\cup l\succ B$, but since $B'\cup j\subseteq \check{B}$ and
$B'\cup j\in\PkNe$, it follows that $B'\cup
j\prec B$, a contradiction.
\end{proof}

\begin{proposition}\label{lem:1}
Let $\prec$ be a compatible order on $\PkNe$. For any $B\in \PkNe$ such that
$\AB$ is nonempty, putting $\check{B}:=\{i_1,\ldots,i_l\}$ with
$i_1\prec\cdots\prec i_l$, then necessarily there exists $j\in \{1,\ldots,l\}$ such
that $B=\{i_j,\ldots,i_l\}$. 
\end{proposition}
\begin{proof}
  Assume $\check{B}\neq B$, otherwise we have simply $j=1$.  Consider $i_j$ the
  element in $B$ with the lowest index in the list $\{1,\ldots,l\}$. Let us prove
  that all successors $i_{j+1},\ldots,i_l$ are also in $B$. Assume $j<l$
  (otherwise we are done), and suppose that $i_{j'}\not\in B$ for some $j<j'\leq
  l$. Then by compatibility, $B=(B\setminus i_j)\cup i_j\prec (B\setminus
  i_j)\cup i_{j'}$. Since $(B\setminus i_j)\cup i_{j'}\subseteq \check{B}$ and
  $(B\setminus i_j)\cup i_{j'}\in\PkNe$, the converse inequality should hold.
\end{proof}

\begin{proposition}\label{prop:7}
Assume that $\prec$ is strongly compatible. Then for all $B\subseteq N$,
$|B|<k$, $\check{B}=B$.
\end{proposition}
\begin{proof}
By Prop. \ref{prop:2} and Cor. \ref{lem:2}, we know that $\AB$ is a Boolean
  lattice with top element denoted by $\check{B}$. Suppose that $\check{B}\neq
  B$. Then there exists $i\in \check{B}\setminus B$, and $B\cup i\in\AB$. Remark
  that $|B\cup i|\leq k$ and $\mathcal{A}(B\cup i)\ni B\cup i$ by
  Prop. \ref{prop:2} and Cor. \ref{lem:2} again. This contradicts the fact that
  the achievable families form a partition of $\PkNe$ (Prop. \ref{prop:0}).
\end{proof}

\begin{proposition}\label{prop:22}
Let $\prec$ be a strongly compatible order on $\PkNe$, and assume w.l.o.g. that
$1\prec 2\prec\cdots\prec n$. Then the collection $\check{\mathcal{B}}$ of
$\check{B}$'s is given by:
\begin{multline*}
\check{\mathcal{B}} = \Big\{\{1,2,\ldots,l\}\cup \{j_1,\ldots,j_{k-1}\}\mid
l=1,\ldots,n-k+1\\ \text{ and } \{j_1,\ldots,j_{k-1}\}\subseteq\{l+1,\ldots,n\}\Big\}
 \bigcup \Big\{A\subseteq N\mid |A|<k\Big\}.
\end{multline*}
If $\prec$
is compatible, then $\check{\mathcal{B}}$ is a subcollection of the above, where
some subsets of at most $k-1$ elements may be absent.
\end{proposition}
\begin{proof}
From Prop. \ref{prop:7}, we know that $\check{\mathcal{B}}$ contains all
subsets having less than $k$ elements. This proves the right part of
``$\displaystyle{\bigcup}$'' in $\check{\mathcal{B}}$. By Prop. \ref{prop:7}
again, the left part uniquely comes from those $B$'s of exactly $k$
elements. Take such a $B$. From Prop. \ref{lem:1}, we know that $\check{B}$
cannot contain elements ranked after the last one of $B$ in the sequence
$1,2,\ldots,n$. In other words, letting $B:=\{l,j_1,\ldots,j_{k-1}\}$, with $l$
the lowest ranked element, we know that $\check{B}=B'\cup
\{l,j_1,\ldots,j_{k-1}\}$, with all elements of $B'$ ranked before $l$. It
remains to show that necessarily $B'$ contains all elements from 1 to $l$
excluded. Assume $j\not\in B'$, $1\leq j<l$. Then it should exist $K\in
\PkNe$, $j\in K\subseteq \check{B}\cup j$, such that $K\succ B$. Since $|B|=k$,
it cannot be that $K\supseteq B$, so that say $j'\in B$ is not in $K$. Hence we have
$j\prec j'$, and by compatibility, $K=(K\setminus j)\cup j \prec (K\setminus
j)\cup j'$. Now, $(K\setminus j)\cup j'\in\PkNe$ and $(K\setminus j)\cup
j'\subseteq \check{B}$, which entails $(K\setminus j)\cup j'\prec B$, so that
$K\prec B$, a contradiction.

Finally, consider that $\prec$ is only compatible. Then by Cor. \ref{lem:2},
there exists $B\in\PkNe$ such that $\AB=\emptyset$. This implies that there
exist some proper subsets of $B$ in $\PkNe$ ranked after $B$, let us call $K$ the last
ranked such subset. Then $|K|<k$, and $\mathcal{A}(K)\neq \{K\}$ since it
contains at least $B$, because all subsets of $B$ are ranked before $K$ by
definition of $K$. Hence $K$ does not belong to $\check{\mathcal{B}}$.  
\end{proof}

We finish this section by explaining why achievable families induced by orders
on $\PkNe$ are generalizations of maximal chains induced by
permutations. Taking
$k=1$, $\mathcal{P}^1_*(N)=N$, and total orders on singletons coincide with
permutations on $N$. Trivially, any order on $N$ is strongly compatible, so
that all achievable families are nonempty lattices. Denoting by $\sigma$ the
permutation corresponding to $\prec$, i.e., $\sigma(1)\prec
\sigma(2)\prec\cdots\prec \sigma(n)$, then
\[
\mathcal{A}(\{\sigma(j)\})=[\{\sigma(j)\},\{\sigma(1),\ldots,\sigma(j)\}], 
\]
i.e., the top element $\check{\{\sigma(j)\}}$ is
$\{\sigma(1),\ldots,\sigma(j)\}$. Then the collection of all top elements
$\check{\{\sigma(j)\}}$ is exactly the maximal chain associated to $\sigma$.

\section{Vertices of $\corek(v)$ induced by achievable families}\label{sec:vert}
Let us consider a game $v$ and its $k$-additive core $\corek(v)$. We suppose
hereafter that $\corek(v)\neq\emptyset$, which is always true for a sufficiently
high $k$. Indeed, taking at worst $k=n$, $v\in\core^n(v)$ always holds.

\subsection{General facts}
A $k$-additive game $v^*$ with M\"obius transform $m^*$ belongs to $\corek(v)$
if and only if it satisfies the system
\begin{align}
\sum_{\substack{K\subseteq A\\ |K|\leq k}}m^*(K) &\geq \sum_{K\subseteq A}m(K),
\quad A\in 2^N\setminus \{\emptyset,N\} \label{eq:8}\\
\sum_{\substack{K\subseteq N\\ |K|\leq k}}m^*(K) &= v(N).\label{eq:9}
\end{align}
The number of variables is $N(k):=\binom{n}{1}+\cdots +\binom{n}{k}$, but due to
(\ref{eq:9}), this gives rise to a ($N(k)-1)$-dim closed
polyhedron. (\ref{eq:8}) is a system of $2^n-2$ inequalities.  The polyhedron is
convex since the convex combination of any two elements of the core is still in
the core, but it is not bounded in general. To see this, consider the simple
following example.
\begin{example}
Consider $n=3$, $k=2$, and a game $v$ defined by its M\"obius transform $m$ with
$m(i)=0.1$, $m(ij)=0.2$ for all $i,j\in N$, and $m(N)=0.1$. Then the system of
inequalities defining the 2-additive core is:
\begin{align*}
m^*(1) & \geq 0.1\\
m^*(2) & \geq 0.1\\
m^*(3) & \geq 0.1\\
m^*(1) + m^*(2) + m^*(12) &\geq 0.4\\
m^*(1) + m^*(3) + m^*(13) &\geq 0.4\\
m^*(2) + m^*(3) + m^*(23) &\geq 0.4\\
m^*(1) + m^*(2) + m^*(3) + m^*(12) + m^*(13) + m^*(23) & = 1.
\end{align*}
Let us write for convenience $m^*:=(m^*(1),m^*(2),m^*(3), m^*(12), m^*(13), m^*(23))$.
Clearly $m^*_0:=(0.2,\;\; 0.1,\;\; 0.1,\;\; 0.2,\;\; 0.2,\;\; 0.2)$ is a solution, as
well as $m^*_0 + t(1,\;\; 0,\;\; 0,\;\; -1,\;\; 0,\;\; 0)$ for any $t\geq 0$. Hence
$(1,\;\; 0,\;\; 0,\;\; -1,\;\; 0,\;\; 0)$ is a ray, and the core is unbounded.
\end{example}
For the monotone core, from Prop. \ref{prop:chj} (i) there is an additional
system of $n2^{n-1}$ inequalities
\begin{equation}\label{eq:10}
\sum_{\substack{K\in[i,L]\\|K|\leq k}}m^*(K) \geq 0, \quad \forall i\in N,
\forall L\ni i.
\end{equation}
For monotone games, Miranda and Grabisch
\cite{migr99a} have proved that the M\"obius transform is bounded as follows:
\[
-\binom{|A|-1}{l'_{|A|}}v(N)\leq m(A)\leq \binom{|A|-1}{l_{|A|}}v(N),\quad
 \forall A\subseteq N,
\]
where $l_{|A|},l'_{|A|}$ are given by:
\begin{enumerate}
\renewcommand{\labelenumi}{(\roman{enumi})}
\item
$\displaystyle{ l_{|A|}=\frac{|A|}{2}} $, and $\displaystyle{ l'_{|A|}=\frac{|A|}{2}-1}$ if $ |A| \equiv 0 (\mathrm{mod}\quad 4)$
\item  
$\displaystyle{ l_{|A|}=\frac{|A|-1}{2}} $, and $\displaystyle{ l'_{|A|}=\frac{|A|-3}{2}} $ or $\displaystyle{
l'_{|A|}=\frac{|A|+1}{2}}$  if $ |A|\equiv 1(\mathrm{mod}\quad 4)$
\item
$\displaystyle{ l_{|A|}=\frac{|A|}{2}-1} $, and $\displaystyle{ l'_{|A|}=\frac{|A|}{2}} $  if $ |A|\equiv 2(\mathrm{mod}\quad 4)$
\item
$\displaystyle{ l_{|A|}=\frac{|A|-3}{2}} $ or $\displaystyle{l_{|A|} =
\frac{|A|+1}{2}}$, and $\displaystyle{  l'_{|A|}=\frac{|A|-1}{2}} $  if $ |A| \equiv 3(\mathrm{mod}\quad 4)$.
\end{enumerate}
Since $v(N)$ is fixed and bounded, the monotone $k$-additive core is always bounded.

\medskip

For $\corek_\infty(v)$, using Prop. \ref{prop:chj} (ii) system (\ref{eq:10}) is
replaced by a system of $N(k)-n$ inequalities:
\begin{equation}\label{eq:11}
m^*(K)\geq 0,\quad K\in \PkNe, |K|>1.
\end{equation}
Since in addition we have $m^*(\{i\})\geq m(\{i\})$, $i\in N$ coming from
(\ref{eq:8}), $m^*$ is bounded from below. Then (\ref{eq:9}) forces $m^*$ to be
bounded from above, so that $\corek_\infty(v)$ is bounded. 

In summary, we have the following.
\begin{proposition}\label{prop:19}
For any game $v$, $\corek(v)$, $\mcorek(v)$ and $\corek_\infty(v)$ are closed
convex\linebreak ($N(k)-1)$-dimensional polyhedra. Only $\mcorek(v)$ and
$\corek_\infty(v)$ are always bounded.
\end{proposition}

The following result about rays of $\corek(v)$ is worthwile to be noted.
\begin{proposition}
The components of rays of $\corek(v)$ do not depend on $v$, but only on $k$ and
$n$. 
\end{proposition}
\begin{proof}
For any polyhedron defined by a system of $m$ inequalities and $n$ variables
(including slack variables) $\mathbf{A}\mathbf{x}=\mathbf{b}$, it is well known
that its conical part is given by $\mathbf{A}\mathbf{x}=\mathbf{0}$, and that
rays (also called basic feasible directions) are particular solutions of the
latter system with $n-m$ non basic components all equal to zero but one (see,
e.g., \cite{chv83}). Hence, components of rays do not depend on $\mathbf{b}$.

Applied to our case, this means that components of rays do not depend on $v$,
but only on $k$ and $n$.
\end{proof}

\subsection{A Shapley-Ichiishi-like result}
We turn now to the characterization of vertices induced by achievable families. 
Let $v$ be a game on $N$, $m$ its M\"obius transform, and $\prec$ be a total
order on $\PkNe$. We define a $k$-additive game $v_\prec$ by its M\"obius
transform as follows:
\begin{equation}\label{eq:1}
m_{\prec}(B) := \begin{cases}
  \sum_{A\in \AB}m(A), &\text{if }  \AB\neq \emptyset\\
  0, &\text{else}
  \end{cases}
\end{equation}
for all $B\in \PkNe$, and $m_\prec(B):=0$ if $B\not\in\PkNe$.

Due to Prop. \ref{prop:0}, $m_\prec$ satisfies $\sum_{B\subseteq
  N}m_\prec(B)=\sum_{B\subseteq N}m(B) = v(N)$, hence $v_{\prec}(N)=v(N)$.

This definition is a generalization of the definition of $\phi^\sigma$ or
$\phi^C$ (see Sec. \ref{sec:prel}). Indeed, denoting by $\sigma$ the permutation on
$N$ corresponding to $\prec$, we get:
\begin{eqnarray*}
m_\prec(\{\sigma(i)\})
 &=& \sum_{A\subseteq\{\sigma(1),\ldots,\sigma(i-1)\}
}m(A\cup\sigma(i))\\ 
 & =& \sum_{A\subseteq\{\sigma(1),\ldots,\sigma(i)\}}m(A) -
\sum_{A\subseteq\{\sigma(1),\ldots,\sigma(i-1)\}}m(A) \\
 & =& v(\{\sigma(1),\ldots,\sigma(i)\}) - v(\{\sigma(1),\ldots,\sigma(i-1)\}) =
 \phi^\sigma(\{\sigma(i)\}) = m^\sigma(\{\sigma(i)\}),
\end{eqnarray*}
where $m^\sigma$ is the M\"obius transform of $\phi^\sigma$ (see
Sec. \ref{sec:prel}).
\begin{proposition}\label{prop:17}
Assume that $\AB$ is a nonempty lattice. Then $v_\prec(\check{B})=v(\check{B})$
if and only if $\{\mathcal{A}(C)\mid C\in\PkNe,C\subseteq
\check{B},\mathcal{A}(C)\neq\emptyset\}$ is a partition of $\mathcal{P}(\check{B})\setminus\{\emptyset\}$.
\end{proposition}
\begin{proof}
We have by Eq. (\ref{eq:1})
\begin{equation}\label{eq:6}
v_\prec(\check{B}) = \sum_{\substack{C\subseteq
    \check{B}\\C\in\PkNe\\\mathcal{A}(C)\neq\emptyset}}m_\prec(C) =
    \sum_{\substack{C\subseteq
    \check{B}\\C\in\PkNe\\\mathcal{A}(C)\neq\emptyset}}\sum_{K\in\mathcal{A}(C)}  
    m(K).
\end{equation}
On the other hand, $v(\check{B})=\sum_{K\subseteq \check{B}}m(K)$. To ensure
$v_\prec(\check{B})=v(\check{B})$ for any $v$, every $K\subseteq
\check{B}$ must appear exactly once in the last sum of (\ref{eq:6}),
which is equivalent to the desired condition. 
\end{proof}

The following is immediate from Prop. \ref{prop:17} and \ref{prop:4}.
\begin{corollary}\label{cor:2}
Assume $\prec$ is compatible, and consider a nonempty achievable family
$\AB$. Then $v_\prec(\check{B})=v(\check{B})$.
\end{corollary}
\begin{proposition}\label{prop:6}
Let us suppose that all nonempty achievable families are lattices. Then $v$
$k$-monotone implies that $v_{\prec}$ is infinitely monotone.
\end{proposition}
\begin{proof}
It remains to show that $m_\prec(B)\geq 0$ for any $B$ such that $1<|B|\leq k$. For
all such $B$ satisfying $\AB\neq\emptyset$,
\[
m_\prec(B)  = \sum_{A\in \AB}m(A) = \sum_{A\in[B,\check{B}]}m(A).
\]
Since $1<|B|\leq k$, by Prop. \ref{prop:chj}, it follows
from $k$-monotonicity that $m_\prec(B)\geq 0$ for all $B\in\PkNe$.
\end{proof}

The next corollary follows from Prop. \ref{prop:2}.
\begin{corollary}\label{cor:4}
Let us suppose that $\prec$ is compatible. Then $v$
$k$-monotone implies that $v_{\prec}$ is infinitely monotone.
\end{corollary}
\begin{theorem}\label{prop:8}
$v$ is $(k+1)$-monotone if and only if for all compatible orders $\prec$,
$v_\prec(A)\geq v(A)$, $\forall A\subseteq N$.
\end{theorem}
\begin{proof}
For any compatible order $\prec$, and any $A\subseteq N$, $A\neq\emptyset$, by
compatibility and Prop. \ref{prop:2}, we can write
\begin{equation}\label{eq:2}
v_\prec(A) = \sum_{\substack{B\subseteq A\\ B\in\PkNe\\
\AB\neq\emptyset}}\sum_{C\in[B, \check{B}]}m(C).
\end{equation}
Let $C\subseteq A$. Then by Prop. \ref{prop:0}, $C\in\AB$ for some $B\subseteq
A$. Indeed $B\subseteq C\subseteq A$. Hence (\ref{eq:2}) writes
\begin{equation}\label{eq:2a}
v_\prec(A) =v(A) + \sum_{\substack{B\subseteq A\\ B\in\PkNe\\
\AB\neq\emptyset}}\sum_{\substack{C\in[B,\check{B}]\\ C\not\subseteq A}}m(C).
\end{equation}

$(\Rightarrow)$ Let us take any compatible order $\prec$. By (\ref{eq:2a}), it
suffices to show that 
\begin{equation}\label{eq:3}
\sum_{\substack{C\in[B, \check{B}]\\ C\not\subseteq A}}m(C)\geq
0,\quad \forall B\subseteq A,B\in\PkNe,\AB\neq\emptyset.
\end{equation}
For simplicity define $\mathcal{C}$ as the set of subsets $C$ satisfying the
condition in the summation in (\ref{eq:3}).  If $\check{B}\subseteq A$, then
$\mathcal{C}=\emptyset$, and so (\ref{eq:3}) holds for such $B$'s. Assume then
that $\check{B}\setminus A\neq\emptyset$. Let us take $i\in \check{B}\setminus
A$. Then $C_0:=B\cup i$ is a minimal element of $\mathcal{C}$, of cardinality
$1<|B|+1\leq k+1$. Observe that $[C_0,\check{B}]\subseteq \mathcal{C}$, and that
it is a Boolean sublattice of $[B,\check{B}]$. Hence, $(k+1)$-monotonicity
implies that $\sum_{C\in [C_0,\check{B}]}m(C)\geq 0$ (see Prop. \ref{prop:chj}).

Consider $j\in\check{B}\setminus A$, $j\neq i$. If no such $j$ exists, then
$[C_0,\check{B}]=\mathcal{C}$, and we have shown (\ref{eq:3}) for such $B$'s.
Otherwise, define $C_1:=B\cup j$ and the interval $[C_1,\check{B}\setminus i]$,
which is disjoint from $[C_0,\check{B}]$. Applying again $(k+1)$-monotonicity we
deduce that $\sum_{D\in [C_1,\check{B}]}m(D)\geq 0$. Continuing
this process until all elements of $\check{B}\setminus A$ have been taken, the
set $\mathcal{C}$ has been partitioned into intervals $[B\cup
i,\check{B}],[B\cup j,\check{B}\setminus i],[B\cup k,\check{B}\setminus
\{i,j\}],\ldots, [B\cup l,A\cup l] $ where the sum of $m(C)$ over these
intervals is non negative by $(k+1)$-monotonicity. Hence (\ref{eq:3}) holds in
any case and the sufficiency is proved.

$(\Leftarrow)$ Consider $K,L\subseteq N$ such that $1<|K|\leq k+1$ and $L\supseteq
K$. We have to prove that $\sum_{C\in [K,L]}m(C)\geq 0$. Without
loss of generality, let us assume for simplicity that $K:=\{i,i+1,\ldots,l\}$
and $L:=\{1,\ldots,l\}$, with $l-k\leq i< l\leq n$. Define $B:=K\setminus
i=\{i+1,\ldots,l\}$ and $A:=L\setminus i$. Take a total order on $\PkNe$ as
follows:
\begin{enumerate}
\item put first all subsets in $\PkLe$, with increasing cardinality, except $B$
  which is put the last
\item then put remaining subsets in $\PkNe$ such that they form a compatible
 order (for example: consider the above fixed sequence in $\PkLe$ augmented with
 the empty set as first element of the sequence, then take any subset $D$ in
 $N\setminus L$ belonging to $\PkNe$, and add it to any subset of the sequence,
 discarding subsets not in $\PkNe$. Do this for any subset $D$ of $N\setminus
 L$).
\item subsets in $\PkLe$ with same cardinality are ordered according to the
lexicographic order, which means in particular $1\prec 2\prec\cdots\prec l$.
\end{enumerate}
One can check that such an order is compatible\footnote{For example, with $n=5,
l=4, i=3, k=3$:
\[
1\prec 2\prec 3\prec 12\prec13\prec14\prec23\prec24\prec34\prec 123\prec
124\prec 134\prec 234\prec 4\prec 5\prec 51\prec 52\cdots
\]}. 
By construction, we have $\AB=[B,L]$. Indeed, for any $C\in\AB$, any subset of
$C$ in $\PkNe$ is ranked before $B$. Moreover, $[K,L]=[B\cup i,L]=\{C\in\AB\mid
C\not\subseteq A\}$. Now, take any $B'\neq B$ in $\PkLe$ such that $B'\subseteq
A$. Let us prove that any $C\in\mathcal{A}(B')$ is such that $i\not\in C$,
or equivalently $C\subseteq A$. Indeed, up to the fact that $B$ is ranked last,
the sequence $\PkLe$ forms a strongly compatible order. Adapting
slightly Prop. \ref{prop:7}, it is easy to see that if $|B'|<k$, then either
$\check{B}'=B'$ or $\mathcal{A}(B')=\emptyset$, the latter arising if $B'\supset
B$. Then trivially any $C\in\mathcal{A}(B')$ satisfies $C\subseteq A$. Assume now
$|B'|=k$. If $B'$ contains some $j\prec i$, then $B'\cup i$ cannot belong to
$\mathcal{A}(B')$ since by lexicographic ordering $B'\cup i\setminus j$ is
ranked after $B'$, which implies that for any $C\in\mathcal{A}(B')$, $i\not\in
C$. Hence, the condition $i\in C$ can be true for some $C\in\mathcal{A}(B')$
only if all elements of $B'$ are ranked after $i$. But since
$B=\{i+1,\ldots,l\}$, this implies that either $B'=B$, a contradiction, or $B'$
does not exist (if $|B|<k$).

Let us apply the dominance condition for $v_\prec(A)$. Using (\ref{eq:2a}),
dominance is equivalent to write:
\[
\sum_{\substack{B\subseteq A\\ B\in\PkNe\\
\AB\neq\emptyset}}\sum_{\substack{C\in [B,\check{B}]\\ C\not\subseteq A}}m(C)\geq 0.
\]
Using the above, this sum reduces to $\sum_{C\in[K, L]}m(C)\geq
0$. This finishes the proof.
\end{proof}

The following is an interesting property of the system \{(\ref{eq:8}), (\ref{eq:9})\}.
\begin{proposition}\label{prop:19a}
Let $\prec$ be a compatible order. Then the linear system of equalities
$v_\prec(\check{B})=v(\check{B})$, for all $\check{B}$'s induced by $\prec$,
is triangular with no zero on the diagonal, and hence has a unique solution. 
\end{proposition}
\begin{proof}
We consider w.l.o.g. that $1\prec 2\prec\cdots\prec n$ and consider the binary
order $\prec^2$ for ordering variables $m^*(B)$.

Delete all variables such that $\mathcal{A}(B)=\emptyset$, and consider the list
of subsets in $\PkNe$ corresponding to non deleted variables. Take all $B$'s in
the list, and their corresponding $\check{B}$'s (always exist since by
compatibility, $\mathcal{A}(B)$ is a lattice). They are all different by
Prop. \ref{prop:0}, so we get a linear system of the same number of equations
(namely $v_\prec(\check{B})=v(\check{B})$) and variables. Take one particular
equation corresponding to $B$. Then variables used in this equation are necessarily
$m^*(B)$ itself (because $\check{B}\supseteq B$), and some variables ranked
before $B$ in the binary order. Indeed, if $\check{B}=B$, then all variables
used in the equation are ranked before $B$ by $\prec^2$. If $\check{B}\neq B$,
supersets $B'$ of $B$ in $\PkNe$ are ranked after $B$ by $\prec^2$ (because
$\prec^2$ is $\subseteq$-compatible), and ranked before $B$ by $\prec$
(otherwise $\mathcal{A}(B)$ would not contain $\check{B}$), but since they
contain $B$, necessarily $\mathcal{A}(B')=\emptyset$, so corresponding variables
are deleted.

Hence the system is triangular.    
\end{proof}

Note that the proof holds under the condition that all achievable families
are lattices, so compatibility is even not necessary.

\begin{theorem}\label{prop:20}
Let $v$ be a $(k+1)$-monotone game. Then
\begin{enumerate}
\item If $\prec$ is strongly compatible, then $v_\prec$ is a vertex of
  $\corek(v)$. 
\item If $\prec$ is compatible, then $v_\prec$ is a vertex of
  $\corek_\infty(v)$.
\end{enumerate} 
\end{theorem}
\begin{proof}
By standard results on polyhedra, it suffices to show that $v_\prec$ is an
element of $\corek(v)$ (resp. $\corek_\infty(v)$) satisfying at least $N(k)-1$
linearly independent equalities among (\ref{eq:8}) (resp. among (\ref{eq:8}) and
(\ref{eq:11})). Assume $\prec$ is compatible. Then by Cor. \ref{cor:4},
$v_\prec$ is infinitely monotone, and it dominates $v$ by $(k+1)$-monotonicity
(Th. \ref{prop:8}).  Moreover, for any $B\in \PkNe$, $\AB$ is either empty or a
lattice, hence either $m_\prec(B)=0$ or $v_\prec(\check{B})=v(\check{B})$ by
Cor. \ref{cor:2}. Since if $|B|=1$, $\AB\neq\emptyset$, this gives $N(k)$
equalities in the system defining $\corek_\infty(v)$, including (\ref{eq:9}),
hence we have the exact number of equalities required, which form a nonsingular
system by Prop \ref{prop:19a}, and (ii) is proved. If the order is strongly
compatible, then all achievable families are lattices, which proves the result
for $\corek(v)$, since again by Prop. \ref{prop:19a}, the system is nonsingular.
\end{proof}

%(see alternative proofs in appendix) 
\begin{remark}
Vertices induced by (strongly) compatible orders are also vertices of the
monotone $k$-additive core. They are induced only by dominance constraints, not
by monotonicity constraints. 
\end{remark}
\begin{remark} Cor. \ref{cor:2} generalizes
Prop. \ref{prop:mave}, while Theorems \ref{prop:8} and \ref{prop:20} generalize
the Shapley-Ichiishi results summarized in Th. \ref{th:sha}. Indeed, recall that
convexity is 2-monotonicity. Then clearly Th. \ref{prop:8} is a generalization
of (i) $\Rightarrow$ (ii) of Th. \ref{th:sha}, and Th. \ref{prop:20} (i) is a
part of (iv) in Th. \ref{th:sha}. But as it will become clear below, all
vertices are not recovered by achievable families, mainly because they can
induce only infinitely monotone games. In particular, $\mcorek(v)$ contains many
more vertices.
\end{remark}

Let us examine more precisely the number of vertices induced by strongly
compatible orders. In fact, there are much fewer than expected, since many
strongly compatible orders lead to the same $v_\prec$. The following is a
consequence of Prop. \ref{prop:22}.
\begin{corollary}\label{cor:6}
The number of vertices of $\corek(v)$ given by strongly compatible orders is at
most $\frac{n!}{k!}$.
\end{corollary}
\begin{proof}
Given the order $1\prec 2\prec\cdots\prec n$, a permutation over the last $k$
singletons would not change the collection $\check{\mathcal{B}}$. 
\end{proof}

Note that when $k=1$, we recover the fact that vertices are induced by all
permutations, and that with $k=n$, we find only one vertex (which is in fact the
only vertex of $\core^n(v)$), which is $v$ itself (use Prop. \ref{prop:22} and
the definition of $m_\prec$).

\subsection{Other vertices}
In this last section we give some insights about other vertices. Even for the
(non monotonic) $k$-additive core, in general for $k\neq 1,n$, not all vertices
are induced by strongly compatible orders. However, for the case $k=n-1$, it is
possible to find all vertices of $\corek(v)$. For $1<k<n-1$ and also for the
monotonic core, the problem becomes highly combinatorial.
\begin{theorem}
Let $v$ be any game in $\mathcal{G}(N)$, with M\"obius transform $m$. 
\begin{enumerate}
\item If $m(N)>0$, $\core^{n-1}(v)$ contains exactly $2^{n-1}$ (if $n$ is even) or
$2^{n-1}-1$ (if $n$ is odd) vertices, among which $n$ vertices come from strongly
compatible orders. They are given by their M\"obius transform:
\[
m_{B_0}^*(K) = \begin{cases}
  m(K), & \text{if } K\not\supseteq B_0\\
  m(K)+(-1)^{|K\setminus B_0|}m(N), & \text{else}
  \end{cases}
\] 
for all $B_0\subset N$ such that $|N\setminus B_0|$ is odd. 
\item If $m(N)=0$, then there is only one vertex, which is
$v$ itself.
\item If $m(N)<0$, $\core^{n-1}(v)$ contains exactly $2^{n-1}-1$ (if $n$ is odd)
  or $2^{n-1}-2$ (if $n$ is even) vertices, of which none comes from a strongly
  compatible order. They are given by their M\"obius transform:
\[
m_{B_0}^*(K) = \begin{cases}
  m(K), & \text{if } K\not\supseteq B_0\\
  m(K)-(-1)^{|K\setminus B_0|}m(N), & \text{else}
  \end{cases}
\] 
for all $B_0\subset N$ such that $|N\setminus B_0|$ is even. 
\end{enumerate}
\end{theorem}
\begin{proof}
We assume $m(N)\geq 0$ (the proof is much the same for the case $m(N)\leq 0$). 
We consider the system of $2^n-1$ inequalities \{(\ref{eq:8}),(\ref{eq:9})\}, which
has $N(n-1)=2^n-2$ variables. We have to fix $2^n-2$ equalities, among which
(\ref{eq:9}), so we have to choose only one inequality in (\ref{eq:8}) to remain
strict, say for $B_0\subset  N$, $B_0\neq\emptyset$:
\begin{equation}\label{eq:p1}
\sum_{K\subseteq B_0} m^*(K) > \sum_{K\subseteq B_0} m(K).
\end{equation}
From the definition of the M\"obius transform, we have
\[
0 = m^*(N) = \sum_{K\subseteq N}(-1)^{|N\setminus K|}v^*(K).
\]
Note that for any $\emptyset\neq K\subseteq N$, $v^*(K)$ is the left member of some
inequality or equality of the system. Hence, by turning all inequalities into
equalities,  we get, by doing the above summation on the system,
 $0 = m(N)$. Hence, if
$m(N)=0$, there is only one vertex, which is $v$ itself, otherwise taking
equality everywhere gives a system with no solution. Since strict inequality
holds only for $B_0\subset N$, we get instead $0>m(N)$ if $|N\setminus B_0|$ is even,
and $0<m(N)$ if $|N\setminus B_0|$ is odd. The first case is impossible by
assumption on $m(N)$, so only the case where $|N\setminus B_0|$ odd can produce a
vertex. Note that if $|B_0|=n-1$, we recover all $n$ vertices induced by
strongly compatible orders. In total we get $\binom{n}{n-1} + \binom{n}{n-3} +
\cdots +\binom{n}{1} = 2^{n-1}$ potential different vertices when $n$ is even, and
$2^{n-1}-1$ when $n$ is odd. Clearly, there is no other possibility.

It remains to show that the corresponding system of equalities is non singular,
and eventually to solve it. Assume $B_0\subset N$ in (\ref{eq:p1}) is chosen. 
From the linear system of equalities we easily deduce $m^*(K)=m(K)$ for
all $K\not\supseteq B_0$. Substituting into all equations, the system reduces to
\begin{align*}
\sum_{K\subseteq B\setminus B_0} m^*(B_0\cup K) & = \sum_{K\subseteq B\setminus
  B_0} m(B_0\cup K), \quad \forall B\supset B_0, B\neq N\\
\sum_{K\subset N\setminus B_0} m^*(B_0\cup K) & = \sum_{K\subset N\setminus
  B_0} m(B_0\cup K) + m(N).
\end{align*}
$B_0$ being present everywhere, we may rename all variables after deleting
$B_0$, i.e., we set $N'=N\setminus B_0$, $m'(A):=m(A\cup B_0)$ and
$m'^*(A)=m^*(A\cup B_0)$, for all $A\subseteq N'$. The system becomes
\begin{align*}
\sum_{K\subseteq B} m'^*(K) & = \sum_{K\subseteq B} m'(K), \quad \forall B\subset
N'\\
\sum_{K\subset N'} m'^*(K) & = \sum_{K\subset N'} m'(K) + m'(N').
\end{align*}
Summing equations of the system as above, i.e., computing $\sum_{B\subseteq
  N'}(-1)^{|N'\setminus B|}\sum_{K\subseteq B}m'^*(K)$, we get
  $m'^*(\emptyset)=m'(\emptyset) + m'(N')$, or equivalently $m^*(B_0)= m(B_0)
  +m(N)$. Substituting in the above system, we get a system which is triangular
  (use, e.g., Prop. \ref{prop:19a} with $k=n=n'$). We get easily $m^*(K)=m(K)
  +(-1)^{|K\setminus B_0|}m(N)$, for all $K\supseteq B_0$.
\end{proof}

\bibliographystyle{plain}
\bibliography{../BIB/fuzzy,../BIB/grabisch,../BIB/general}

\end{document}